\documentclass[a4paper,10pt]{article}
\usepackage[T1]{fontenc}
\usepackage[cp1250]{inputenc}
\usepackage[polish,english]{babel}
\usepackage[cp1250]{inputenc}
\usepackage[numbers]{natbib}
\usepackage{graphicx}

\title{Analysis of Norms Game in networked societies}
\author{Pawel Sobkowicz}
\date{19th October 2003}

\begin{document}

\maketitle
\section*{Abstract}
Norms, defined as generally accepted behaviour in societies without central authority (and thus distinguished from laws), are very powerful mechanism leading to coherent behaviour of the society members. This paper examines, within a simple numerical simulation, the various effects that may lead to norm formation and stability. The approach has been first used by Axelrod, who proposed two step model of norm and meta-norm enforcement. We present here an extension and detailed analysis of the original work, as well as several new ideas that may bear on the norm establishment mechanisms in societies. It turns  out that a relatively simple model for simulated norm enforcement predicts persistent norm breaking even when it is associated with high punishment levels. The key factors appear to be the combination of the level of penalty for breaking the norm and proximity of norm enforcers. We also study a totally different mechanism of norm establishment, without meta-norms but using instead the direct bonus mechanism to norm-enforcers.

\section{Definitions of the {\sc Norms Game} variants}
The original work of \citet{axelrod86-1}, later reprinted and expanded, as Chapter 3 of \cite{axelrod97-1} has described a possible scenario for modelling norm formation, called the {\sc Norms Game} (NG), and its higher level variant, the {\sc Meta Norms Game} (MNG). The games are based on extended Prisoner's Dilemma game, including additional action by the rest of community:  \emph{punishment} of a \emph{defector} (NG), and additional possibility of punishment of \emph{for failing to punish the defector} (MNG).  The idea behind the model is to see if such enforcement may lead to formation of stable cooperative society, and whether simple enforcement has to be augmented by meta-enforcement, to achieve stability of the system. 

\subsection{The classic formulation of Norms Game.} 
 The model presented by Axelrod is based on an $N$-person PD game. Within the game, each player $i$ has several "action opportunities" each turn, during which he can either \textbf{play safe}, in which case he gets no benefit, or  \textbf{defect} in which case he gets a \textbf{treasure} of $T > 0$, while players $j \neq i$ get extra \textbf{harm} $H < 0$. The willingness of an agent $i$ to commit a defection is called \textbf{boldness} $b_i$. For obvious reasons, the only stable strategy in such game is that of total defection --- each "safe player" is at distinct disadvantage with the defectors.

The next step is the introduction of \textbf{norm enforcement} activities (the norm being understood as the neutral, safe behaviour). 

If the player $i$ does not defect, other players do not take action. If, on the other hand, $i$ does break the norm, other players $j$ may, with probability $v_j$ (\textbf{vengefulness}), punish him for misbehaviour. In the original treatment, the \textbf{punishment} value $P$ was assumed greater than the treasure $T$. At the same time, each punishing agent pays punishment costs $E$ (\textbf{enforcement}). The punishment for defection constitutes the proper Norms Game: society uses active measures to enforce the norms. 

After each round of interactions between agents, Axelrod used simple population adjustment, giving higher population ratios to those players with higher overall benefits. He has also used a small "mutation" rate to introduce small random shifts in strategies. Results of simulations were presented after 100 generations. In Axelrod's simulations two out of five runs  ended up in boldness $b \approx 1$ and $v \approx 0$, while other simulations ended with various levels of vengefulness, with very small boldness. The pure norm enforcing behaviour, depending on the simulation history or initial conditions led to either almost fully defecting society or to one pacified by vengefulness. Axelrod's argument was that pure NG is too weak to ensure norm respecting behaviour.

Introducing Meta Norms Game, as an extension of the previous model, Axelrod proposed that if agent $i$ defects and agent $j$ does not punish $i$, then with probability $v'$ any other agent $k$ should punish $j$ \emph{for failing to enforce the norm!} The meta-punishment $P'$ and cost associated with it, $E'$ were assumed to be equal to $P$ and $E$.

More importantly, the \textbf{meta-vengefulness} $v'$ was assumed to be equal to $v$. This decision was based on rather vague arguments. The link definitely simplified calculations, but, as noted by Axelrod himself, prohibited situations when certain agents would use different values of $v$ and $v'$ to their advantage. Results of simulations with $v=v'$ were uniform: in all cases the society converged to almost pure high vengefulness, low boldness, norm-observing one.

From his simulations Axelrod has concluded that, for pure Norms Game, \begin{quote} \ldots at first, boldness levels fell dramatically due to the vengefulness in the population. Then, gradually, the amount of vengefulness also fell because there was no direct incentive to pay the enforcement cost of punishing the defection. Once vengefulness became rare, the average level of boldness rose again, and then the norm completely collapsed. Moreover, the collapse was a stable outcome. \end{quote}

In contrast,  the inclusion of meta-norm enforcement had provided the agents with a strong incentive to increase their vengefulness and this to promote the norm (decrease boldness). This was indeed confirmed by the simulation results. 

From one point of view this is obvious consequence of the model, in which some characteristics are penalised (boldness in case of NG, boldness and lack of vengefulness in the case of MNG). Lack meta-vengefulness is not penalised at all, just the opposite --- it carries its own  cost, and is thus disadvantageous evolutionarily. This leads to  minimization of the meta-vengefulness in the evolution process -- unless, as in Axelrod's model, it is forcibly fixed, for example by coupling it to vengefulness. By setting $v=v'$, $P=P'$ and $E=E'$ Axelrod has not only simplified the simulations, but also made a sort of self-policing model. He has thus excluded  analysis  treating meta-enforcement and enforcement as independent yet interacting phenomena.

Additionally, because in the original form, the Norms Game was played with a very small number of agents (20) no scaling effects were visible. One of such effects is the cumulation of penalties  due to the model in which \textit{anyone} is capable of punishing everyone, and individual punishments add up. This results in penalties easily "overcoming" benefits of trespassing. Despite the mentioned oversimplifications of the simulations themselves, the MNG model seems to be interesting enough to use it as a basis for more detailed studies. A significant and open question remains whether the results would be changed by larger and longer simulations and by more flexible model.

\subsection{Modified form of Norms Game with single enforcers} 

To extend the validity of the model and to check if the assumptions made by Axelrod could be derived from more general principles we propose a modified form of MNG. 

We assume that $N_T$ agents ($N_T \gg 1$) are connected via a network of connections, each interacting with a number of neighbors. In different simulations we have used four basic network topologies, typical for biological or social networks. These types included: {\bf fully connected network}, in which every agent is connected with all others; {\bf random network}, where agents are connected by randomly distributed links; {\bf nearest neighbour} network, where the links are highly clustered and  finally the {\bf small world} networks. The three latter types of networks may be called {\bf local}, as the number of agents that a given agent interacts with (actively or passively) is much smaller than the total number of agents. This limits the influence and sets stage for possible {\emph indirect} influences.
  General description of such networks may be found in \citet{albert02-1, dorogovtsev02-1, dorogovtsev02-2, dorogovtsev02-3, newman00-1, newman03-1, newman03-3}. All these networks are characterized by filling ratio, which relates to the average number of links per node (agent). The larger the filling ratio, the more agents are directly connected. In our simulations we have used ratios from 0.005 to 0.02 (which corresponds to circa 5 to 20 links per agent. The difference between the various types of local networks are in the way the agents are connected. 

In random (RAND) networks the links are distributed randomly, thus we have a meshed network of links, with agents differing in number of the neighbours, and no general structure. 

The nearest neighbor (NN) nets are formed by linking certain number of closest neighbors together. The easiest way to visualize such network is to place the agents on an imaginary circle and connecting each agent to $n$ neighbors. The number of connections per agent is constant. Interesting property of NN networks is that for small filling factors, agents on the opposite points at the circle to communicate must go through many intermediaries. For 2000 agents and number of neighbors set at 10, the longest `distance' is 100 `hops'. This suggests influence on the changes in norm adherence and enforcement: any change of behavior of the agent $i$ is seen immediately by his closest neighbors (who can see him through direct links) but only after considerable delay and filtering by members of society located far from $i$.  

The Small World (SW) networks, introduced and popularized in recent years, reproduce a curious fact observed in many natural and human-produced networks, namely that the distance between any two nodes of the network, measured in links needed to connect them, is usually much smaller than that in  nearest neighbor or even random networks of the same filling ratio. The name of the network category comes exactly from such observation. One of the models for SW networks is a simple reworking of the NN model: one takes a small (even very small!) fraction of the links from between nearest neighbors and applies them instead between random agents. Keeping in mind the visualization of NN networks as nodes along a circle, this corresponds to adding connections that criss-cross the circle at random. Due to such shortcuts, even if their number is very small, the average distance between any two nodes drops dramatically. Thus we have a network that for each agent, locally is very similar to NN model (as most of the neighbors are, in fact, the same), but globally the communication through the network is much  faster.

Within our  MNG scenario each agent has a choice between acting in accordance with a norm, in which case he gets no extra bonus, or breaking the norm (trespassing), in which case he gets extra bonus (treasure) $T$. If the agent $i$ misbehaves, every agent in the population is harmed by this action. We can imagine this corresponds, for example, to the agent stealing some `communal property'. The treasure $T$ is the benefit to the trespasser, but the harm $H$ to the community may be greater than $T$. It is then assumed that the harm $H$ is  divided evenly among population, each agent `losing' $H/N_T$. 

Let's denote the probability of misbehavior (boldness) of agent $i$ by $b_i$. The payoff $X(i)$ of agent $i$ is (without any punishments) given by
\begin{equation}
 X(i) = b_i T - N_T {\langle b \rangle}_T H / N_T = b_i T - H {\langle b \rangle}_T 
\end{equation}
where ${\langle b \rangle}_T$ denotes average of $b_i$ over the total population. With $H>T$ the overall average payoff becomes negative ${\langle b \rangle}_T (T-H)$ and is \textit{disadvantageous} to the community, but as any agent with boldness higher than the average gets more benefit than the rest, such behaviour gives evolutionary advantage, and thus multiplies. Natural end of such a process is a free-for-all, $b_i = 1$ society.

As in Axelrod's work we propose that to curb boldness, society enforces the norm through penalization of trespassing. Our model differs slightly in the way the process of detection and punishment is encoded. 

Each agent watches his neighbours for acts of  misbehavior and is may be willing to punish for such acts. For an agent $j$ with vengefulness $v_j$ the probability of detection that another agent $i$ has broken the norm is just $v_j$. The watchfulness and vengefulness result in enforcement effort $E$, the overall cost of being vigilant to an agent $j$ being $v_j E$.  The first agent to notice that someone has broken the norm punishes the transgressor with a punishment of $P$.  Thus only one punishment act take place and there is no cumulation effects, but the value of $P$ may be set freely. To calculate the expected penalty value we start with the probability of no-one detecting the misbehaviour of $i$ is
\begin{equation}
\prod_{(j\neq i)} (1-v_j),
\end{equation}
where the product is taken over all agents $j$ with \textbf{links} to $i$, that is  over the  \textbf{neighbours} of $i$. 
As a result the penalty for agent $i$ would be given by
\begin{equation}
\pi_P(i) = \left( 1 - \prod_{(j\neq i)} (1-v_j) \right) b_i P,
\label{Ppunish1}
\end{equation}
that is proportional to probability is at least one agent detecting the act of breaking the norm times probability of such act times the penalty value. Without meta-norms the payoff for player $i$ is thus
\begin{equation}
b_i T - H {\langle b \rangle}_T - v_i E - \left( 1 - \prod_{(j\neq i)} (1-v_j) \right) b_i P 
\end{equation}

So far the difference between the modified model and Axelrod's one was in the fact that the punishment takes place only once. The real change comes in the way the meta-norm is introduced. In the original model \textit{every failure to punish a norm breaker was meta-punished}. Here,  we have only one punishing agent (the first one to detect the trespasser). The problem  is how should the meta-enforcer decide whether non-punishment by certain agent was intentional (due to lack of vengefulness or it's small value) or just a result of coming in second, after the enforcement has already taken place? 
One way is to set the probability of meta-punishment as being proportional to the lack of vengefulness of the observed agent: 
\begin{equation}
X_{MP}(j,i) = v'_j (1-v_i), 
\end{equation} 
with $v'_j$ being the meta-vengefulness of the observing agent $j$ and $1-v_i$ being the lack of vengefulness of the observed agent $i$. Thus the expected value of meta-punishment for an agent $i$ is
\begin{equation}
\pi_{MP}(i) = \left( 1 - \prod_{(j\neq i)} \left( 1- X_{MP}(j,i) \right) \right) P'
\label{MPpunish1}
\end{equation}

As our simulations were conceived as extension of the work of Axelrod, we have used similar convention of allowing discretized values of $b_i$, $v_i$ and $v'_i$ (from 0 to 1 with steps of 0.1). To see if such discretization does not produce any artificial effects we have also performed calculations in which boldness, vengefulness and meta-vengefulness were allowed to take arbitrary values between 0 and 1. To our slight surprise the results did not differ significantly. The specific results presented in this paper refer to the discrete model.

\subsubsection{Results of NG and MNG for large fully connected networks}
Lets turn now to the NG and MNG played in relatively large, fully connected networks. Here every agent interacts with all agents as in the case of the small model of Axelrod, and the evolution seems rather simple. The most important fact here is that even a single enforcer or meta-enforcer (say an accidental mutant) would be able to influence the payoffs of all agents, and direct the course of evolution of population. 

Lets consider first the situation where there is no meta-norm enforcement at all ($P'=0$). If one starts from a random population in all simulations the final average value of vengefulness was $v=0$. This is due to additional costs borne by the enforcers, and the pressure to minimize this cost. However, as initially there were some norm enforcers (and remember that in a fully connected network just one mutant would suffice to enforce the norm on the whole population!) the enforcement put a clear divide between two basic situations. When the treasure $T$ is greater than  $P$ it pays to be bold, and $b\rightarrow 1$. On the contrary, for $T<P$ we have $b\rightarrow 0$. This corresponds to common sense observation that the norm would be obeyed if the penalty for breaking it would be sure (which is made probable by large number of observers) and greater than the benefits.

When the meta-norm enforcement is present the situation shifts to meta-level. In all simulations the final average value of meta-vengefulness goes to zero ($v'\rightarrow 0$) --- due to the same arguments as vengefulness dropped in previously described simulation. However, as already noted, in a fully connected network even a single meta-enforcer is able to meta-punish everyone in society! Thus, although nominally the number of these meta-enforcers goes is zero in the stable situation, their initial influence and occasional mutant presence are sufficient to distinguish the situation from the previously described case of $P'=0$. 

The results of simulations differ in small details (such as the number of iterations it takes to achieve stability) but in principle are governed by two simple rules.
\begin{itemize}
\item The boldness is decided by relation of $T$ vs. $P$, as described above
\item The vengefulness is decided by relation of the cost of vengefulness $E$ and the meta-punishment $P'$
\end{itemize}  

Again, the above results are very much common sense. Decoupling $v$ from $v'$ results in $v'\rightarrow 0$ for MNG, which contrasts the assumed $v=v'$ in Axelrod approach. Large size of the population leads to elimination of `undecided' runs --- all simulations ended with uniform populations.

\subsubsection{Local networks}
In contrast to the fully connected case, the \textbf{local networks} present much more interesting norms dependence on the payoff parameters. 

As in the original formulation of MNG, in addition to a breeding mechanism, in which certain percentage of agents with the best payoffs were allowed to breed (at the cost of the worst faring agents), we have used additional mutation process. This was simulated as allowing certain fraction of agents (given by mutation ratio) to change their characteristic $b_i$, $v_i$ and $v'_i$ values between interactions. Although the number of mutants per iteration was usually kept very small in relation with the population size (e.g. 2 mutants out of 2000 per iteration), its effects were quite important. Without mutation populations sometimes froze in a given configuration, `far' from results obtained with even minimal mutation rates.

Another effect we set out to investigate was the dependence of the behaviour on the starting conditions.  Two general types of starting conditions were used: in the first the initial populations had randomly distributed values of $b_i, v_i$ and $v'_i$. In the second, we have started from \textit{pure} society of, for example, $b_i=0$, $v_i=1$ and $v'_i=0$, or $b_i=1$, $v_i=1$ and $v'_i=0$  ---  to observe if such pure societies were evolutionary stable with respect to small mutation rates. It turned out that regardless of the starting condition, if the mutation was present, the final state was very similar.

We present general results of simulation runs through a few examples, and discuss the role of the starting conditions. The simulations were run for 2000 agents using fixed values of $P=20$, $P'=3$, $E'=1$, $H=12$, with top 1\% of the population being allowed to reproduce each iteration (at the expense of the bottom 1\%). Other parameters ($T$, $E$) were used to control the simulations and the dependence on them is presented and discussed below.

\subsubsection{Similarities and differences between random, NN and SW networks}

In our simulations we  compare results for all three types of local networks. Despite the fact that the three types of locally connected networks have many different properties (see, for example \cite{albert02-1, dorogovtsev02-1, dorogovtsev02-2, dorogovtsev02-3}) the results of our simulations for random, small world and nearest neighbour networks are strikingly similar (see Figure 1). It turns out that the key factor is the common characteristic of all models, given in the first approximation  by the average number of neighbours. The number of neighbors is related network filling factor $\phi$, defined as ratio of the actual links in the network to the total possible number of links:
\begin{equation}
\phi = \frac{N_{LINKS}}{N_{T}(N_{T}-1)/2}.
\end{equation}
For regular networks, such as NN network, the number of neighbors is the same for all agents, $N_N=N_{LINKS}=\phi N_{TOT} /2$. For small world networks, constructed as in this work, the $N_N$ defined above is a very close measure of the number of neighbors, with very few exceptions. For random networks, the dispersion in number of links per agent is greater, but still $N_N$ gives the average value. 

The differences between local network topologies for the same filling factor values are relatively small. Thus the use of the word `local' is justified: behaviour of an agent is then determined by his immediate surrounding, mainly by the number of agents who observe it and vice-versa, the number of agents it observes.

The two vertical columns of plots in Figure 1 correspond to two cases of $E/P'$ ratio. The left column has $E/P' = 2/3 <1 $, which in the case of fully connected network would correspond to large vengefulness $v$ -- so, to strong pressure to decrease $b$. The right column has $E/P' = 5/3$, with meta-penalty smaller than the cost of being vigilant there is little incentive to be a norm enforcer. Indeed, the results of simulations for the two cases are strikingly different. For $E/P'<1$ $v$ is relatively large and $b$ resembles the step-like function found in fully connected network. $v'$ remains relatively small. On the other hand, for $E/P' > 1$ $v$ is smaller, $v'$ larger, and the step like character of $b$ is replaced by gradual growth. This corresponds to the presence of trespassers who, in local environments, find that locally the effective punishment is smaller than the treasure value. Thus even at $T/P$ much smaller than 1 we have relatively high average $b$.  

\subsubsection{Discussion of the dependence on the filling factor}

Figure 2 presents another set of comparisons of the average final $b$, $v$ and $v'$ values, this time for a single type of the network topology (Small World) but differing by the filling factor $\phi$. The top row corresponds to $\phi=0.005$, middle row to $\phi=0.01$, bottom row to $\phi=0.02$. Corresponding average number of neighbours changes from 5 to 20. For the case of $E/P'=5/3$ (right column) the difference is visible mostly in lowering values of $v$ and $v'$ below $T/P =1$, and in more pronounced kink in $b$ at $T/P=1$ with increasing $\phi$.

For $E/P' = 2/3$ one can observe two effects:
\begin{itemize}
\item a shift of the step-like increase of $b$ as function of $T/P$ to values lower than 1;
\item an increase in vengefulness $v$ in the region of $T$ far below $P$ for $E=2$, where simple explanations suggest no dependence of $v$ over $T$.
\end{itemize}

The first effect can find a qualitative explanation through analogy with simplified, small size fully connected model.
Then the finite size of neighbourhood changes the condition for the misbehaviour to be advantageous from $T>P$ to:
\begin{equation}
T_i > P \left( 1- \prod_{j\neq i}(1-v_j)\right) \simeq P \left( 1- (1-v)^{N_N} \right)
\end{equation}
with $v$ being the average vengefulness. The expression $P \left( 1- (1-v)^{N_N} \right)$ may be dubbed the effective deterrent, as it weights the penalty with probability of being punished. For large number of neighbors $N_N$ the term $(1-v)^{N_N}$   goes to zero, and we recover the $T>P$ condition. For small $N_N$ the difference is visible as the shift of the step in $b(T)$ function.

\subsection{Simulations with group enforcement}
The previous sections have discussed the extreme opposite to Axelrod's assumption of additive punishment --- only a single agent was executing the punishment, and it was assumed that this single agent had enough power to effect the punishment on anyone it had links with. Within this model a single enforcer or meta-enforcer could `police' the whole neighbourhood.

It is interesting to see if the results obtained in previous simulations would be changed when one would require some greater number of agents, acting in unison, to effect a punishment or meta-punishment. Such situation of `group action', or coalition forming against a trespasser seems more similar to `real life'. 

Lets assume that the punishment would take place if at least $k_P$ neighbors of trespassing agent $i$ would be willing to enforce the norm. Similarly, meta-punishment would require $k_{MP}$ neighbors to act together. In our calculations we would limit ourselves to low values of $k_P$ and $k_{MP}$ of 2 and 3.

To obtain the expressions for group punishment model one starts with $\Phi_m(i)$ and $\Psi_m(i)$ defined as: probability that exactly $m$ neighbors if agent $i$ would be willing to enforce the norm (or meta-norm) on $i$. We have:
\begin{eqnarray}
\Phi_0(i) & = & \prod_{j\neq i} (1-v_j) \\ 
\Phi_1(i) & = & \sum_{k\neq i} \quad \prod_{j\neq i,k} v_k (1-v_j) \\
\Phi_2(i) & = & \sum_{k\neq i}  \sum_{l>k, l\neq i}  v_k v_l \prod_{j\neq i,k,l} (1-v_j)\\
\Phi_3(i) & = & \sum_{k\neq i}  \sum_{l>k, l\neq i} \quad  \sum_{m>l, m\neq i}  v_k v_l v_m\prod_{j\neq i,k,l,m}
 (1-v_j)
\end{eqnarray}

\begin{eqnarray}
\Psi_0(i) & = & \prod_{j\neq i} \left(1-v'_j(1-v_i)\right) \\ 
\Psi_1(i) & = & \sum_{k\neq i} v'_k  (1-v_i) \prod_{j\neq i,k} \left(1-v'_j (1-v_i) \right) \\
\Psi_2(i) & = & \sum_{k\neq i} \sum_{l>k, l\neq i} v'_k v'_l (1-v_i)^2  \prod_{j\neq i,k,l} \left(1-v'_j (1-v_i) \right) \\
\Psi_3(i) & = & \sum_{k\neq i} \sum_{l>k, l\neq i} \quad \sum_{m>l, m\neq i} v'_k v'_l v'_m (1-v_i)^3  \prod_{j\neq i,k,l,m}  \left( 1-v'_j (1-v_i) \right) 
\end{eqnarray}

Thus, requiring that more than $k_{P}$ agents (who are neighbors) decide group together and punish transgressor $i$ (or respectively $k_{MP}$ agents group to meta-punish $i$) we have the expression for punishment (meta-punishment)
\begin{eqnarray}
\pi_{P}^{k_P}(i) & = & \left( 1 - \sum_{k=0}^{k_P} \Phi_k(i)\right) b_i P  \label{PPunishk} \\
\pi_{MP}^{k_P}(i) & = & \left( 1 - \sum_{k=0}^{k_{MP}} \Psi_k(i) \right)  P' \label{MPpunishk}
\end{eqnarray}
Obviously, for $k_P=0$ and $k_{MP}=0$ (which corresponds to just one agent being able to effect punishment) the above expressions reduce to Equations \ref{Ppunish1}, \ref{MPpunish1}.

Figure 3 compares results for SW network at low filling factor $\phi=0.005$, where the problems related to the need to find enough enforcers for the group action should be most visible, for three values of $k_P = 1, 2, 3$ (top, middle and bottom row respectively). The shift of high $b$ regime to regions of $T/P < 1$ is even more pronounced than in Figure 2, the reason being clear: it is increasingly more difficult to find two and three agents acting in unison in neighborhoods of average size of 5. However, there is no radical change in behaviour of the system as the whole.

\section{To enforce or to break the norm? }

In Axelrod's study as well as in the results presented above, boldness and vengefulness  were treated as separate characteristics of an agent. Details of our simulations show that in most cases the best payoffs are achieved by agents who {\textit at the same time} take advantage from breaking the norm and actively pursue enforcement on other trespassers, that is agents with simultaneously high $b$ and $v$. while perfectly acceptable from mathematical point of view, this is somewhat against psychological expectations: one can either be a policeman or a thief. We dub this case the exclusive model.

To see if such contradiction of choices has influence on the simulated societies we have performed series of runs in which $b_i$ and $v_i$ were coupled through $v_i = 1- b_i$, that is, the more the agent acted as trespasser (higher $b_i$) the smaller was his vengefulness. This simple relation does not reflect the complexity of the issue of the exclusivity of choices, but allows an insight into possible outcomes.

The simulations returned quite interesting results, as presented in Figure 4 for SW network at two filling factors 
($\phi=0.005$ and $\phi=0.02$), with single agent enforcement. For small $T/P$ stable result was $b=0$ and $v=1$. For large $T/P$ the opposite result was true. The width of the transition region depends on the filling factor, for larger number of neighbors the region was much narrower. The change in the value of the  of $E$ resulted in almost direct shift of the of the step-like increase of $b$ as function of $T$ toward lower $T$. The onset of the norm-breaking behaviour was given by simple relation $T_0 = P - E$. This corresponds to condition when the benefit of norm breaking ($T$) exceeds the optimum benefits of the enforcers ($P-E$).

The exclusive $b$--$v$ model could be enhanced in a way that instead of using relation $b=1-v$  (which allows an agent to act `a bit as a policeman and a bit as a thief') one would use more complicated rule. This may be subject of separate investigation.

\section{Norm enforcement through bounty hunting}

Within Axelrod's model vengefulness and meta-vengefulness were directly coupled. In our extension of the approach described above we have decoupled them, and studied influence of the meta-enforcement of the norm persistence. There is, however another interesting possibility, which is based on simplified observations of `real life' situations, where norm enforcement, although in itself costly (through constant watchfulness cost $E$), may be positively reinforced, without resorting to meta-norms. 

The mechanism is quite simple. Suppose the trespasser is caught. Then the agent or agents who have caught him would get a \textbf{bounty} $B$. The bounty may be set arbitrarily, but in out simulations we use the following scenario. The effective penalty $\pi_P^{k_P}(j)$ may be treated as a fine imposed on the trespasser $j$. Part of it accounts for general costs, but the rest, described through fraction $f$, is divided among all agents that have participated in the act of enforcement on the trespasser (or, to formulate it within the model used here, among all agents that \textit{might have participated} in the act of enforcement). This allows for a strategy using \textbf{bounty hunting} as means of increasing the payoff of an agent, and thus increases $v$, which might help in keeping $b$ down.

Suppose we want to establish the payoff for agent $i$ for catching a trespassing agent $j$. The whole bounty $f\pi_P^{k_P}(j)$  has to be divided among all agents $k$ that are neighbors of $j$, in a way that relates to their ability and willingness to enforce the norm. We introduce a simple (though arbitrary) formula that preserves the relation of the individual bounties $B(i,j)$ to the \textit{vengefulness effort}. First we introduce the \textit{summed vengefulness around agent $j$}: 
\begin{equation}
v_j^{S} = \sum_{k}^{(\mbox{\tiny neighbors of } j)} v_k,
\end{equation}
where summation goes over all neighbors of $j$. Using this it is easy to divide the bounty: the bounty for agent $i$ for catching trespasser $j$ (who's willingness to break the norm is given by boldness $b_j$) is
\begin{equation}
B(i,j) = f \pi_P^{k_P}(j) v_i / v_j^S.
\end{equation}
Thus the total bounty payoff for agent $i$ is given by:
\begin{equation}
B(i) = \sum_j B(i,j) = f P v_i \sum_j^{(\mbox{\tiny neighbors of } i)}
\frac{  b_j \sum_{n=0}^{k_P} \left( 1 - \Phi_n(j)\right) }{\sum_{k}^{(\mbox{\tiny neighbors of } j)} v_k }
\end{equation} 
For large number of enforcers with high vengefulness, the bounty would be divided among many agents, and thus relatively small. But suppose there is just a few enforcers --- they would get significant advantage for catching the trespassers. The model has, in place of $P'$ and $E'$ only a single parameter $f$, describing how much of the penalty $P$ is divided among the vengeful agents.

The new formula for payoffs replaces the meta-enforcement terms with the bounty term. The simulation space is now simpler (two-dimensional, with variables $b$ and $v$). Example of simulation results is presented in Figure 5. Left column shows average $b$ and $v$ values as functions of $T/P$, right column shows results for the exclusive model discussed in previous section.

Rows correspond to increasing fractions of the penalty divided as bounty among enforcers, from $f=0.01$, through $f=0.05$ and $f=0.4$ to $f=1$. It is worth noting that as each agent has on average 5 neighbors, these fractions allow, in principle to get quite significant payoff from bounty. With our simulation we have set the enforcement cost at $E=2$ and penalty $P=20$. Thus the bounty from 5 trespassers at $f=0.05$ might reach 5, overcoming the cost of being vengeful.

As a result in the model with independent $b$ and $v$ only for very low $f$ we observe behaviour resembling the one found in MNG simulations. For larger $f$ \textit{both} $b$ and $v$ increase \textit{simultaneously}, in a smooth way. These results form the best illustration of the statement that it pays best to be a thief and policemen at the same time.

For the exclusive case the growth of $b$ is quite similar, with $v$ mirroring it according to condition $v=1-b$. The onset of the increase of $b$ as function of $T/P$ takes place at almost the same values for both the normal and exclusive models

\section{Conclusions}

The conclusions of our models and simulations may be summarized as follows:
\begin{itemize}
\item decoupling norm enforcement and meta-norm enforcement ($v$ and $v'$), set as equal in the original model of Axelrod, has shown that the quantities evolve in very different ways, thus the original assumption was probably introducing artificial effects into results presented in \cite{axelrod86-1,axelrod97-1};
\item for large (more than 20 agents), fully connected societies, boldness and vengefulness followed simple rules, directly relating them to net payoffs for trespassers and enforcers ($T>P$, $P'>E$);  
\item locally networked societies show complicated interplay of various parameters, with the leading role of the $T/P$ and $E/P'$ ratios and strong dependence on the filling factor (number of neighbours). No discernible differences were observed for different network topologies;
\item in all cases we have observed local `pockets' of norm breakers (high $b$ agents) well below the limit of $T/P=1$;
\item a variant of the model, aimed at replicating the dichotomy of choice norm breaking/norm enforcement, in the simplest form of setting $v=1-b$ leads to a smoothed step like behaviour of $b$, with the position of the step decided by $T$ vs $P-E$, and its width by the number of neighbours;
\item a new model, Bounty Norms Game (BNG), alternative to Meta Norms Game, based on the division of part of the penalty among norm enforcers has been presented. Despite totally different nature of the process governing the evolutionary benefits of enforcement the results resemble in general way results from MNG simulations. Simplicity and `real life' background of BNG make it a good candidate for at least some of the norm enforcement studies.
\end{itemize}

\clearpage

%\bibliography{Bibliography}

\clearpage

\section{Figures}

\clearpage

\begin{figure}[!t]
\centering
\includegraphics[height=16cm]{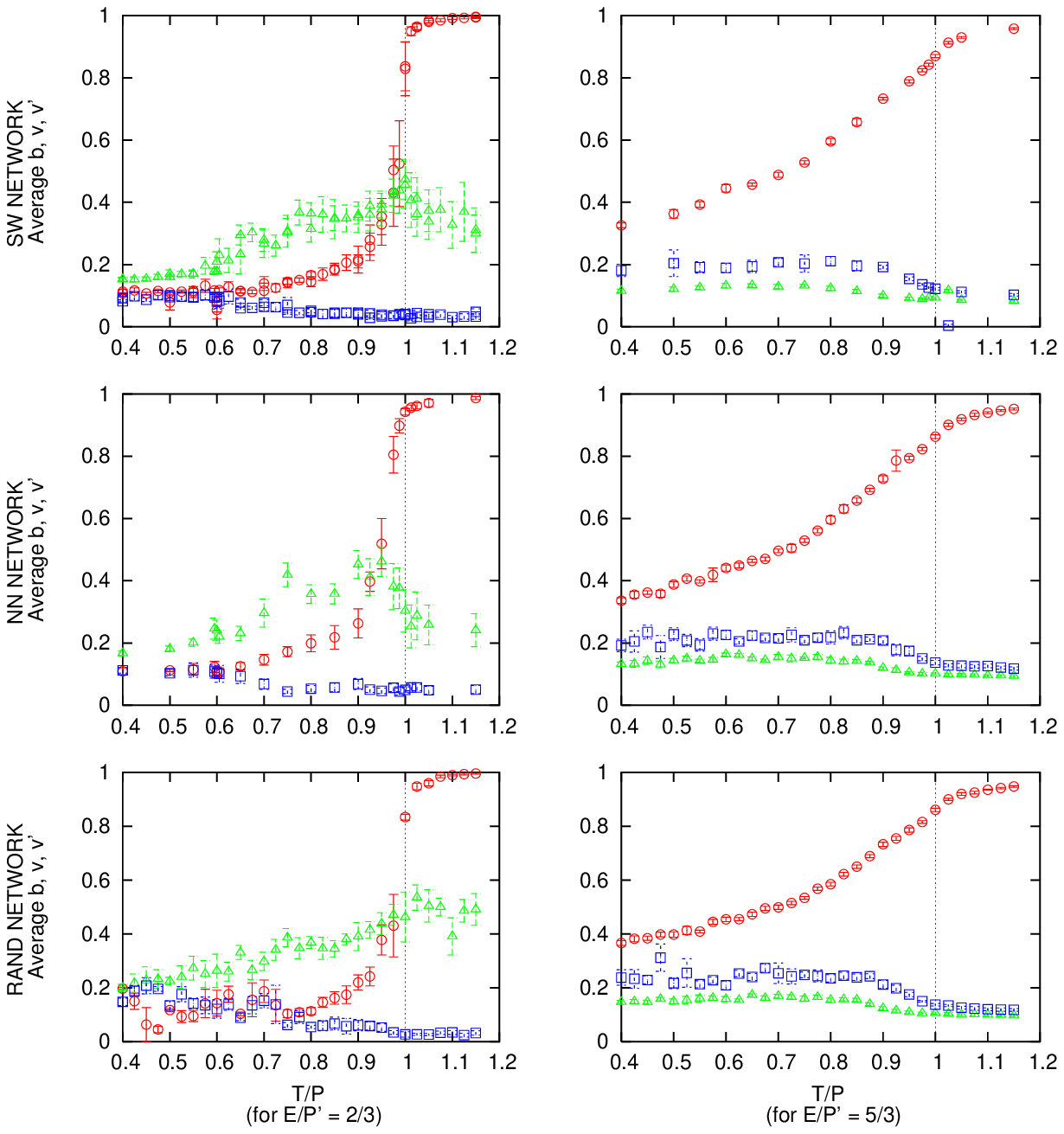}
\caption{Comparison of final values of average $b$, $v$ and $v'$ as functions of the ratio of treasure to penalty ($T/P$) for different network topologies. Top row: SW network, middle row: NN network, bottom row: RAND network. Left column: $E/P'=2/3$, right column $E/P'=5/3$. Red circles -- $b$, green triangles -- $v$, blue squares -- $v'$. Filling factor $\phi = 0.005$.  
\label{figure1}}
\end{figure}

\begin{figure}[!t]
\centering
\includegraphics[height=16cm]{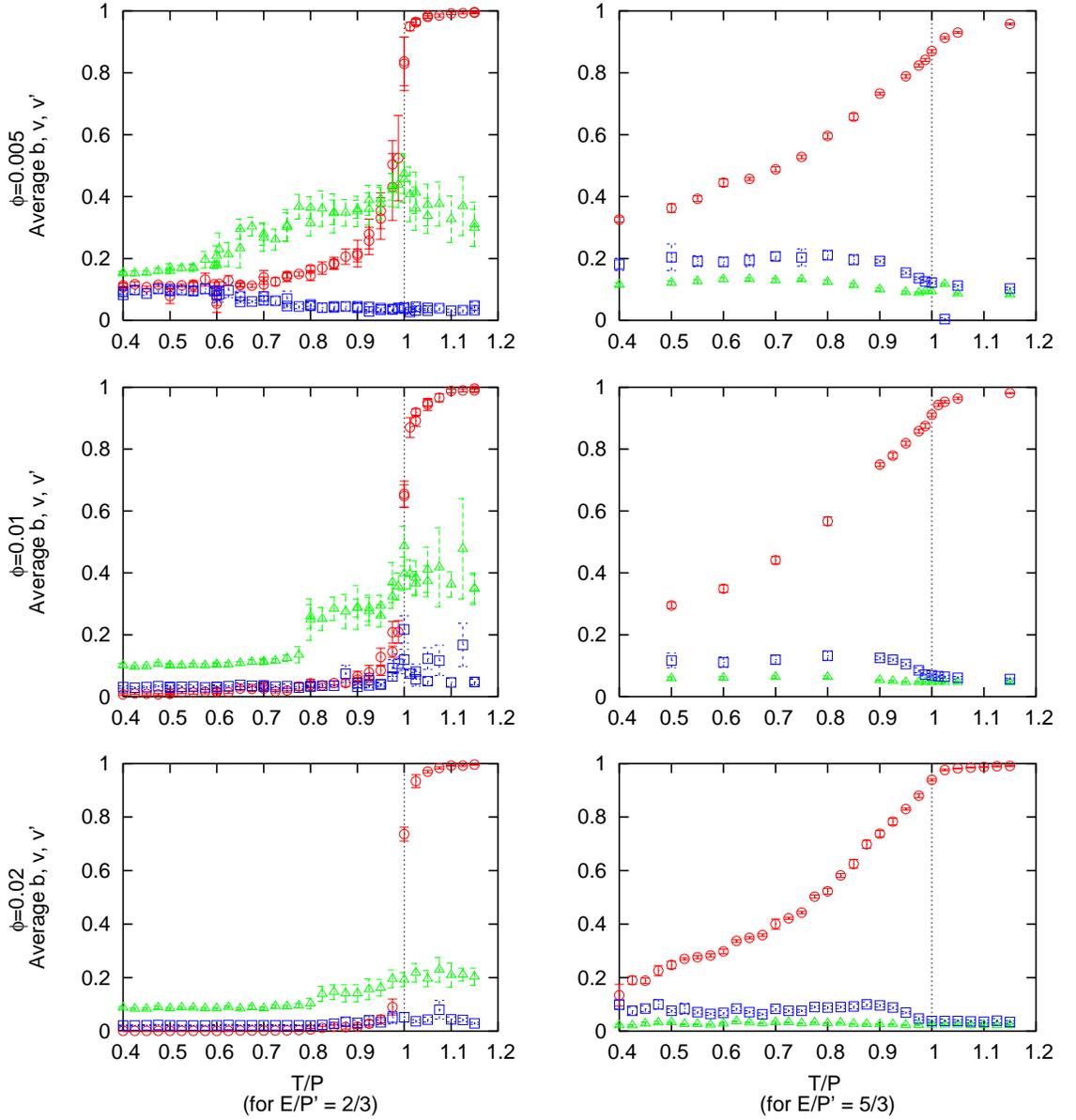}
\caption{Comparison of final values of average $b$, $v$ and $v'$ as functions of the ratio of treasure to penalty ($T/P$) for different filling factors in the case of SW network. Top row: $\phi=0.005$, middle row: $\phi=0.01$, bottom row: $\phi=0.02$. Left column: $E/P'=2/3$, right column $E/P'=5/3$. Red circles -- $b$, green triangles -- $v$, blue squares -- $v'$. 
\label{figure2}}
\end{figure}

\begin{figure}[!t]
\centering
\includegraphics[height=16cm]{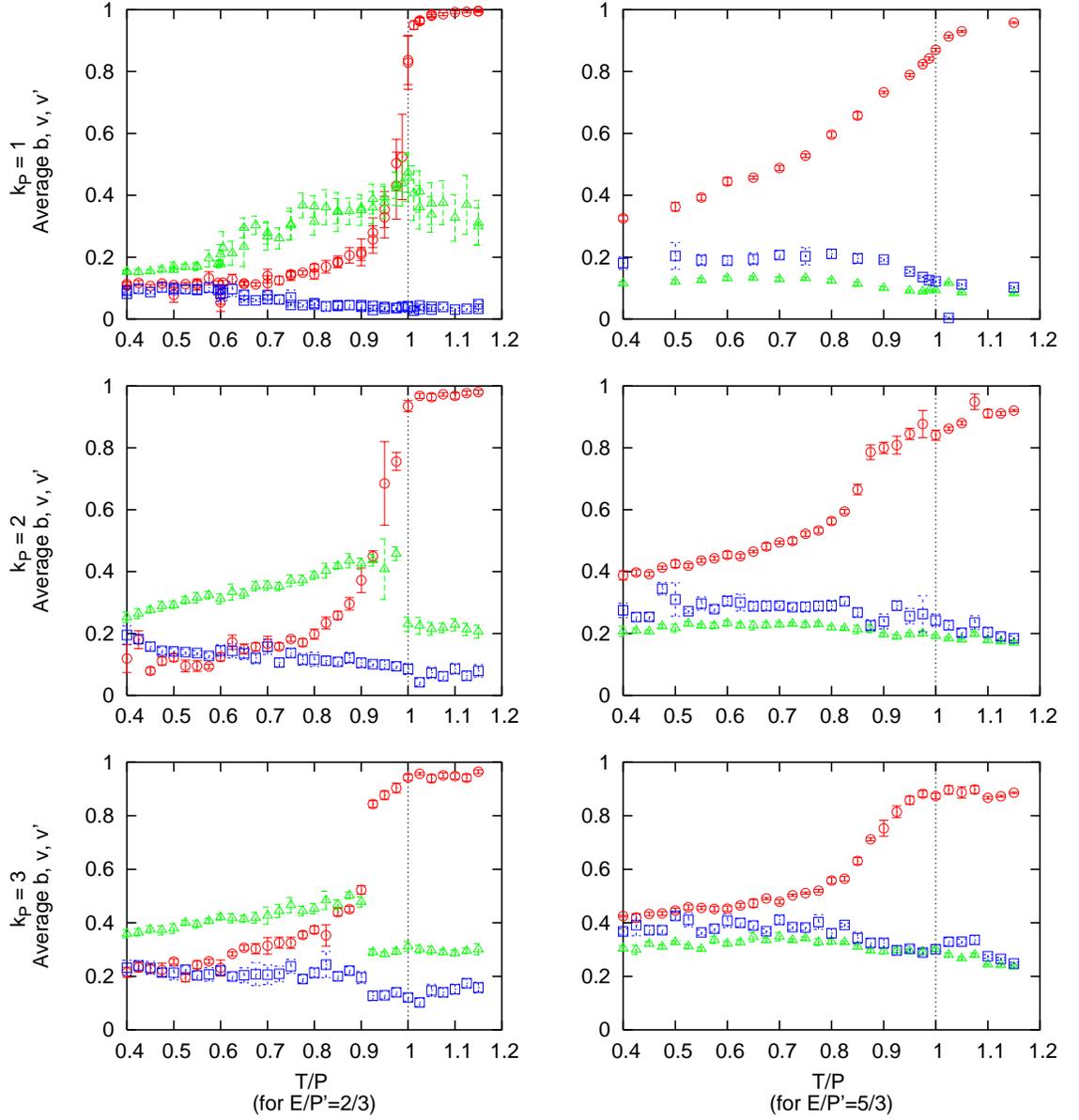}
\caption{Comparison of final values of average $b$, $v$ and $v'$ as functions of the ratio of treasure to penalty ($T/P$) for different sizes of the group needed to enforce the norm, in the case of SW network and $\phi=0.005$. Top row: one agent, middle row: two agents, bottom row: three agents. Left column: $E/P'=2/3$, right column $E/P'=5/3$. Red circles -- $b$, green triangles -- $v$, blue squares -- $v'$. 
\label{figure3}}
\end{figure}

\begin{figure}[!t]
\centering
\includegraphics[height=16cm]{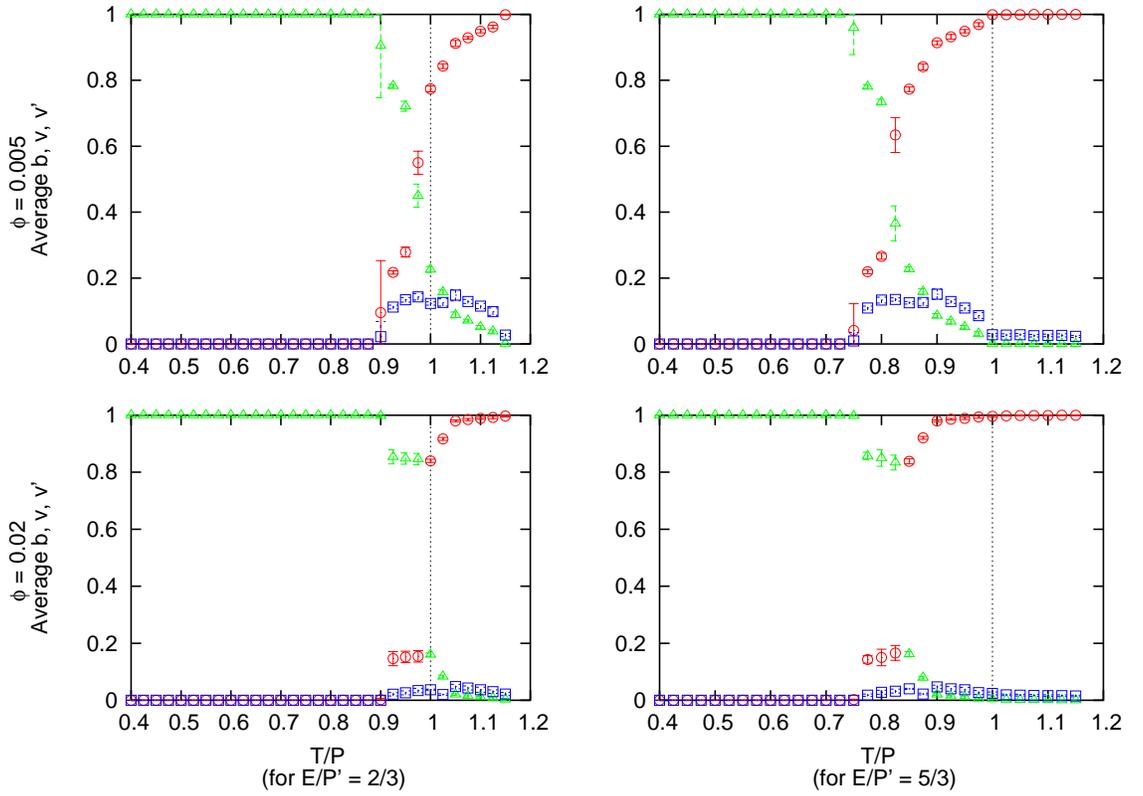}
\caption{Comparison of final values of average $b$, $v$ and $v'$ for the exclusive $b$--$v$ model, in the case of SW network. Top row: $\phi=0.005$, bottom row: $\phi=0.02$. Left column: $E/P'=2/3$, right column $E/P'=5/3$. Red circles -- $b$, green triangles -- $v$, blue squares -- $v'$. 
\label{figure4}}
\end{figure}

\begin{figure}[!t]
\centering
\includegraphics[height=16cm]{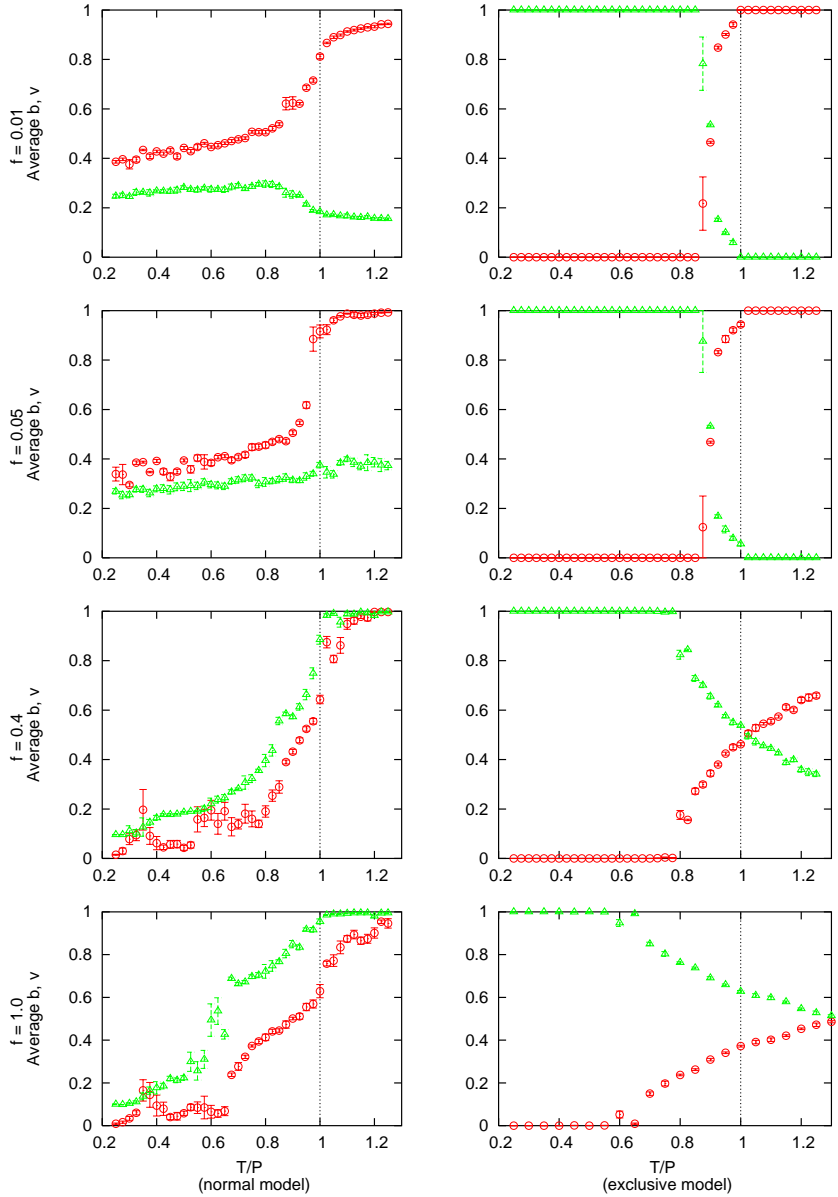}
\caption{Comparison of final values of average $b$, $v$ and $v'$ as functions of the ratio of treasure to penalty ($T/P$) for the bounty hunting model. Top row: $f=0.01$, second row: $f=0.05$, third row: $f=0.4$, bottom row: $f=1$. Left column: model with independent $b$ and $v$; right column: model with exclusive $b$ and $v$ ($v=1-b$). Red circles -- $b$, green triangles -- $v$.
\label{figure5}}
\end{figure}

 \end{document}